\documentclass{iaus}
\usepackage{graphicx}

\title[GRB Progenitors and Observational Criteria] 
{GRB Progenitors and Observational Criteria} 

\author[B. Zhang]   
{Bing Zhang}

\affiliation{
Department of Physics and Astronomy, University of Nevada, Las Vegas, NV 89154, USA 
 email: {\tt zhang@physics.unlv.edu} \\[\affilskip]
}

\pubyear{2012}
\volume{279}  
\pagerange{119--126}
\jname{IAU Symposium 279: Death of Massive Stars: Supernovae \& Gamma-Ray Bursts}
\editors{P. Roming, N. Kawai, E. Pian, eds.}
\begin{document}

\maketitle

\begin{abstract}
Phenomenologically, two classes of GRBs (long/soft vs. short/hard) are
identified based on their $\gamma$-ray properties. The boundary
between the two classes is vague.  Multi-wavelength observations lead
to identification of two types of GRB progenitor: one related to
massive stars (Type II), and another related to compact stars
(Type I). Evidence suggests that the majority of long GRBs belong
to Type II, while at least the majority of nearby short GRBs
belong to Type I. Nonetheless, counter examples do exist. Both
long-duration Type I and short-duration Type II GRBs have been
observed. In this talk, I review the complications in GRB
classification and efforts in diagnosing 
GRB progenitor based on multiple observational criteria. In
particular, I raise the caution to readily accept that all
short/hard GRBs detected by BATSE are due to compact star mergers.
Finally, I propose to introduce ``amplitude'' as the third
dimension (besides ``duration'' and ``hardness'') to quantify burst
properties, and point out that the ``tip-of-iceberg'' effect may 
introduce confusion in defining the physical category of GRBs, 
especially for low-amplitude, high-redshift GRBs.
\end{abstract}

\section{Phenomenological vs. physical classification schemes}

Observations of GRBs with BATSE led to identification of two
phenomenological classes of of GRBs in the duration-hardness
($T_{90} - {\rm HR}$) plane: long/soft vs. short/hard
(Kouveliotou et al. 1993).  The boundary between the two classes
is vague. The duration separation line is around 2 seconds 
in the BATSE band (30 keV - 2 MeV). Long and short GRBs roughly 
comprise 3/4 and 1/4 of the total population.

The main issue of applying the $T_{90}$ criterion to define the
class of a GRB is that $T_{90}$ is detector dependent. GRB pulses
are typically broader at lower energies. Also a more sensitive detector
tends to detect weaker signals which would be otherwise buried in the
noises. Indeed, observations carried out 
with softer detectors such as HETE-2 and {\it Swift} brought confusions to
classification. Among a total 476 GRBs detected by {\it Swift} 
BAT (sensitive in 15 keV - 150 keV) from Dec. 19 2004 to
Dec. 21 2009, only 8\% have $T_{90} < 2$s (Sakamoto et al. 2011), 
much less than the $\sim 1/4$ fraction of the BATSE sample. An 
additional 2\% of {\it Swift} GRBs have a short/hard spike typically 
shorter than or around 2s, but with an extended emission lasting 10's 
to $\sim 100$ seconds.  These ``short GRBs with E.E.'' 
(e.g. Norris \& Bonnell 2006) have $T_{90} \gg 2$s as observed by 
{\it Swift}, but could be ``short'' GRBs if they were detected 
by BATSE. So the unfortunate consequence of the $T_{90}$ classification 
is that the membership to a certain category of the {\em same} GRB 
could change when the detector is changed. 
Nonetheless, the confusion in $T_{90}$ classification 
only arises in the ``grey'' area
between the two classes. For most GRBs, one can still tell
whether they are ``long'' or ``short''.

Follow-up afterglow and host galaxy observations of GRBs led to
the identification of at least two broad categories of progenitor.
Observations led by BeppoSAX, HETE2, and {\it Swift} suggest that 
at least some long GRBs are associated with supernova Type Ic (e.g.
Hjorth et al. 2003; Campana et al. 2006; Pian et al. 2006).
Most long GRB host galaxies are found to be dwarf star-forming 
galaxies (Fruchter et al. 2006). 
These facts establish the connection between 
long GRBs and deaths of massive stars (Woosley 1993).
The breakthrough led by {\it Swift} unveiled that some nearby
short GRBs (or short GRBs with E.E.) have host galaxies that are
elliptical or early-type, with little star formation
(Gehrels et al. 2005; Barthelmy et al. 2005; Berger et al. 2005). 
This points towards another type of 
progenitor that does not involve massive stars, but is likely
related to compact stars, such as NS-NS or NS-BH mergers 
(e.g. Eichler et al. 1992). 

The current observational data suggest that the majority of 
long GRBs belong to the massive star progenitor category
(Type II), and at least the majority of nearby short GRBs
belong to the compact star category (Type I), see e.g. Zhang (2006)
for a discussion on physical classification scheme of GRBs. 
Nonetheless, the cozy picture of ``long GRBs = Type II GRBs'', 
and ``short GRBs = Type I GRBs'' does not always hold. 
Long duration Type I GRBs such as 
GRB 060614 have been observed (e.g. Gehrels et al. 2006; 
Zhang et al. 2007), which show deep upper limits on the 
brightness of any associated supernova, as well as a local galactic
environment with low star formation rate (e.g. Gal-Yam et al. 2006).
Some short-duration
Type II GRBs are also observed, including three highest-redshift, 
``rest-frame short'' (i.e. $T_{90}/(1+z) < 2$ s) GRBs, i.e.
GRB 080913 at $z = 6.7$ (Greiner et al. 2009), GRB 090423 at
$z=8.2$ (Tanvir et al. 2009),
and GRB 090429B at $z=9.4$ (Cucchiara et al. 2011), and one 
observer-frame short GRB 090426 at $z=2.609$ 
(Levesque et al. 2010; Xin et al. 2011).

It is then desirable to answer the following challenging question:

\section{How can one tell the physical class of a GRB based on
the observational data?}

In order to address this question, let's remind ourselves what are
the observational facts that made us believe the existence of 
two distinct classes of progenitor. Zhang et al. (2009) summarized
12 multi-wavelength observational criteria that could be connected
to the physical nature of a GRB, which include
(1) supernova association; (2) specific star formation rate of
the host galaxy; (3) position inside the host galaxy; (4) duration;
(5) hardness; (6) spectral lag; (7) statistical correlations 
(e.g. $E_p-E_{\rm \gamma,iso}$, $E_p - L_{\rm \gamma,iso}$,
$L-{\rm lag}$); (8) energetics and beaming; (9) afterglow 
properties (medium density and spatial profile); (10) redshift 
distribution; (11) luminosity function; and (12) gravitational
wave signal. Except criterion (12), which could be more definitive
but is more difficult to carry out, other criteria have been applied
to the known GRBs.
Criteria (10) and (11) are statistical, which rely on a large sample
of data. Some interesting results have been obtained, which will be
discussed in \S3 below. Other criteria can be applied to
individual GRBs, and the above particular order of the criteria 
is based on how closely a particular criterion is directly relevant
to the progenitor. One can see that ``duration'' and ``hardness'', which
are used in phenomenological classification, are not
direct indications of GRB progenitor. This is not surprising, since
the bimodal distribution has been known since the BATSE era, but
it was after BeppoSAX, HETE, and Swift when people identified
the two broad progenitor types. The first three criteria (SN,
host galaxy, and position within the host galaxy) carry much more
weight in defining the physical category of a GRB.

\begin{figure}[ht]
\begin{center}
 \includegraphics[width=5in]{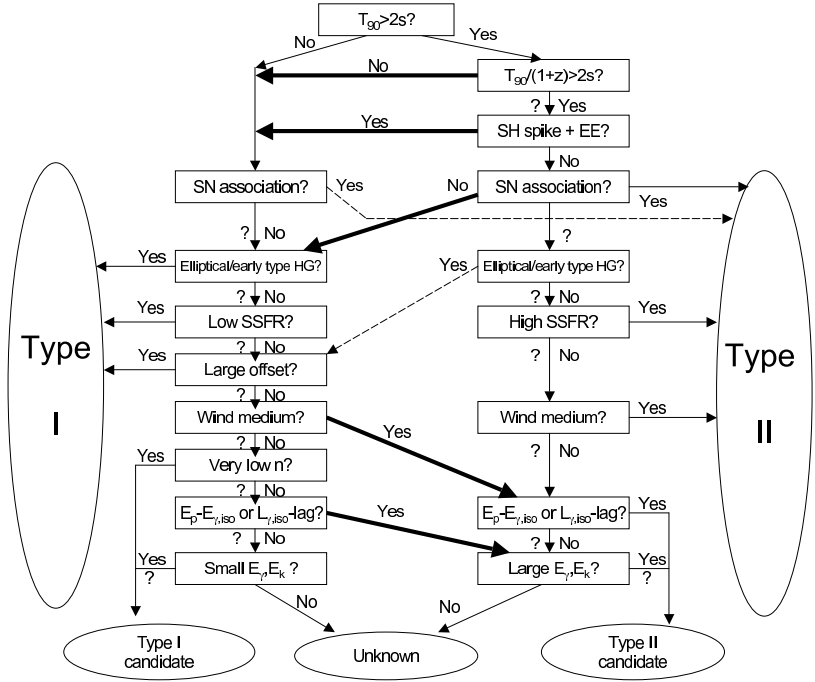} 
 \caption{The flowchart of using multiple observational criteria to 
diagnose the physical category of a GRB. After the initial duration 
criteria (which are used to start with a rough separation), the 
sequence of the applied criteria are based
on their degree of relevance to the progenitor. From Zhang et al. (2009).}
   \label{flowchart}
\end{center}
\end{figure}

A flowchart to diagnose the physical category of a GRB based
on multiple observational criteria was proposed in Zhang et al.
(2009), see Fig.1. Several noticeable features of the flowchart
are the following: 1. Even though duration can be used to roughly
separate GRBs, there are several bridges that allow bursts to
break the duration separation line. Noticeable examples include
the long-duration Type I GRB 060614, rest-frame short high-$z$
GRBs 080913, 090423, and 090429B, as well as observer-frame short
GRB 090426. 2. The upper criteria (SNe and host galaxy properties)
carry more weight, and can directly lead a burst to Type I or
Type II; 3. The lower criteria invoking afterglow modeling or
empirical correlations carry less weight, and would only lead 
bursts to Type I or Type II ``candidates''.

Such a flowchart has been applied to study currently observed
long and short GRBs with afterglow data, and it proves to work 
well (Kann et al.  2010, 2011). 
In order to apply the flowchart, one needs to have a lot of extra
information (afterglow, SN, host galaxy, redshift, etc.) other than 
prompt $\gamma$-ray properties. With $\gamma$-ray information only,
one cannot determine the physical type of GRB with high confidence.
Nonetheless, one could give a ``guess''. This is particularly 
important for a GRB trigger team, since early on there is no
afterglow, redshift, SN, and host information. An alert from the
team would help the follow-up observers to decide how significant
a burst is. The Swift team essentially applies the following 
flowchart to predicts the category of the GRB (Fig.2).

\begin{figure}[ht]
\begin{center}
 \includegraphics[width=3in]{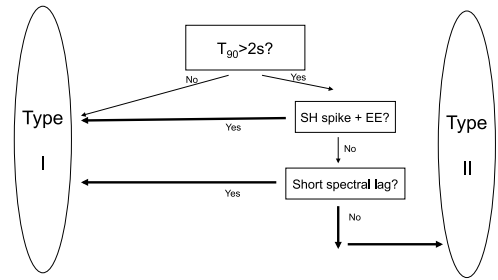} 
 \caption{The flowchart used by the Swift team.  }
   \label{swift}
\end{center}
\end{figure}

Some other efforts have been carried out to use fewer parameters
to identify the physical category of a GRB. For example, L\"u et al.
(2010) showed that for GRBs with $z$ measurements, the parameter 
$\varepsilon \equiv E_{\gamma,iso,52}/E_{p,z,2}^{5/3}$ can be 
a good indicator.  The high-$\varepsilon$ vs.
low-$\varepsilon$ categories are found to be more closely related 
to Type II vs. Type I, respectively, rather than the traditional
long vs. short classification.

\section{Are all short/hard GRBs of a compact star merger origin?}

The leading model for Type I GRBs is NS-NS or NS-BH mergers.
An immediately question would be: are all short/hard
GRBs as detected by BATSE are due to NS-NS or NS-BH mergers?

Most $z$-known short-duration GRBs are predominantly located at 
redshifts below 1 (Berger 2009). The peak luminosity spans in a wide
range (Virgili et al. 2011), from $\sim 7\times 10^{48}~{\rm erg~s^{-1}}$ 
(for GRB 050509B, Gehrels et al. 2005) to 
$\sim 3.8\times 10^{52}~{\rm erg~s^{-1}}$ (for GRB 090510, De Pasquale
et al. 2010). An intuitive response to the above question is ``perhaps
not''. Unlike massive star progenitor that has an extended envelope
that can collimate the jet, a NS-NS merger system does not have a
natural collimator. A broad jet can be launched, but never highly
collimated (e.g. $\sim 30^\circ$, Rezzolla et al. 2011). The 
neutrino annihilation energy power is typically low. Even if the
Blandford-Znajek mechanism is invoked, to produce a burst similar
to GRB 090510 requires far-stretching of the models. Indeed,
Zhang et al. (2009) suggested that high-$z$, high-$L$ short GRB
may not be of an Type I origin. Panaitescu (2011) suggested
a Type II origin of GRB 090510 based on its wind-like density profile
in afterglow modeling.

There are two approaches to address whether all short GRBs are due to
compact star mergers. The first approach
is to use multiple criteria to identify a Gold sample of Type I GRBs,
and check whether it is a good representation of the entire short/hard
GRB population. Zhang et al. (2009) took this approach and concluded
that it may not be. The second approach is to take the 
observed short GRBs as one population, and statistically compare
it with the long GRB population to see the difference. The host galaxy 
study by Berger (2009) and Fong et al. (2010) is of this type. They
show that short GRB host galaxies and the afterglow positions inside
the hosts are indeed statistically different from those of long GRBs.
The sample for such study is still small, and is dominated by the nearby
low-$L$ short GRBs (which are Type I). So the conclusion does
not necessarily support that all BATSE short/hard GRBs are Type I.

\begin{figure}[ht]
\begin{center}
 \includegraphics[width=2.6in]{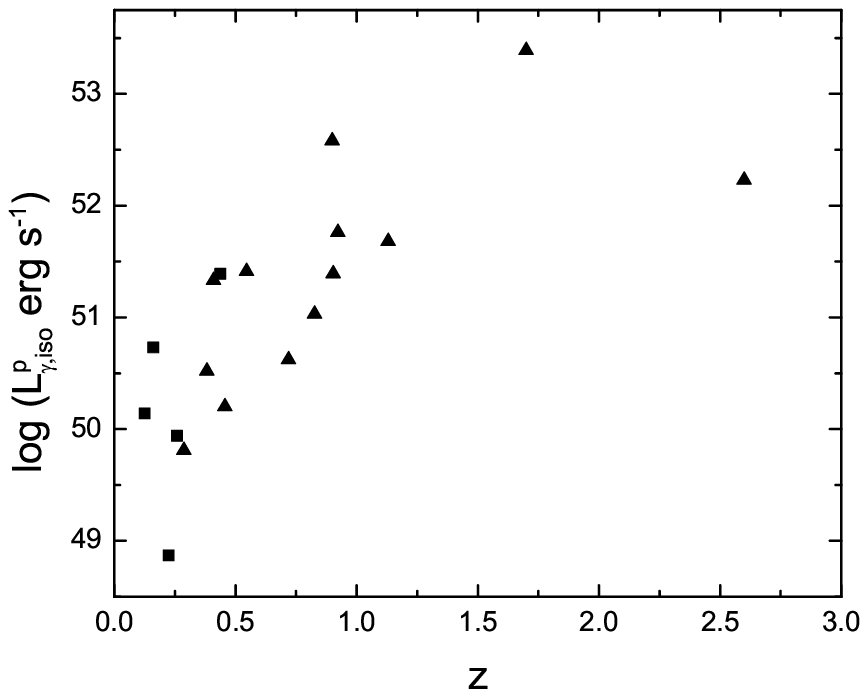} 
 \includegraphics[width=2.6in]{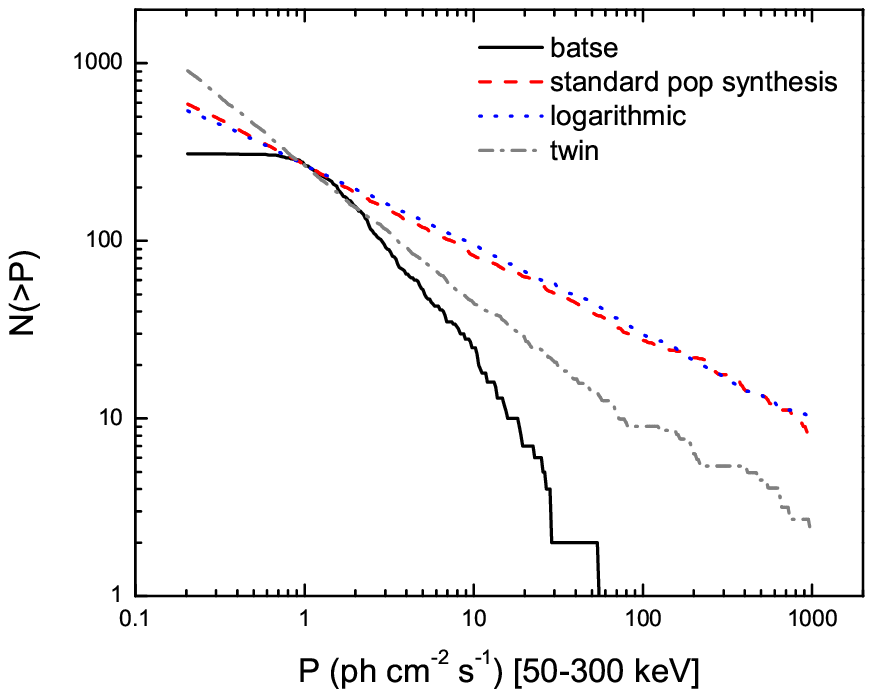} 
 \caption{{\em Left}: the luminosity ($L$) - redshift ($z$) 
distribution of the $z$-known short GRB population detected by
Swift (before June 2009). One can see low-$z$ clustering, and
lack of significant increase of low-$L$ short GRBs with respect
to high-$L$ short GRBs, suggesting a shallow luminosity function;
{\em Right}: Within the compact star merger model, a shallow luminosity
function is translated into a shallow peak flux distribution, which
is inconsistent with the $\log N - \log P$ distribution of the
BATSE short GRBs. This suggests that the compact star merger model
cannot simultaneously interpret both the $z$-known Swift sample
and the $z$-unknown BATSE sample. From Virgili et al. (2011).}
   \label{short}
\end{center}
\end{figure}

In Virgili et al. (2011), we also performed an analysis using the second
approach, but applied the gamma-ray data (instead of the host galaxy
data). We assume that the $z$-known short-duration GRBs detected by 
Swift and the $z$-unknown short/hard GRBs detected by BATSE belong to
the same parent sample, and test whether the NS-NS and NS-BH merger
models can simultaneously reproduce both daughter samples. As shown in
Fig. 3, this is extremely difficult, if not impossible. The reason
is that the $L-z$ distribution of the Swift sample demands a shallow
luminosity function, since one does not see a significant increase of 
the number of low-$L$ short GRBs with respect to high-$L$ short GRBs.
Since merger models predict a low-$z$ clustering due to the merger time
delay with respect to star formation, this shallow luminosity function
is translated to a shallow peak flux distribution ($\log N - \log P$),
which cannot interpret the BATSE short GRB $\log N - \log P$ data.
Notice that even considering the highest fraction of ``prompt mergers''
(e.g. Belczynski et al. 2007), the shallow $\log N - \log P$ cannot 
be removed. There are two possibilities: 1. the Swift sample is not
a good manifestation of the BATSE sample. A good fraction of the BATSE
short GRBs are different, which may be a mix of Type I and Type II GRBs;
2. All short GRBs are Type I, but the NS-NS and NS-BH merger modes are
not the correct one. Other models (e.g. accretion induced collapses)
may be needed. In any case, using the short GRB sample to infer
gravitational wave detection rate (e.g. Coward et al. 2012) 
may be pre-mature.

\section{Introducing ``amplitude'' as the third dimension and the
confusion regimes of GRB classification}

In principle, a short GRB can be the ``tip-of-iceberg'' of a long
GRB, if the extended longer duration emission is not bright enough
to emerge from the detector background. So it is important to introduce
the third dimension, i.e. ``amplitude'', besides ``duration'' and
``hardness'' to classify GRBs. 

One can define a parameter $f \equiv F_p / F_b$, where $F_p$ is the
peak flux, while $F_b$ is the background flux. For a long GRB, if we
ideally scale down the flux so that the measured $T_{90}$ becomes
shorter than 2 seconds, we will have a dimmer peak flux, $F'_p$, 
for which one can define a new parameter $f_{\rm eff} 
\equiv F'_p/F_b$. For short GRBs, since $T_{90} < 2$ s initially,
one has $f_{\rm eff}=f$. For long GRBs, $f_{\rm eff}$ defines a limit
below which the ``tip-of-iceberg'' effect becomes important.

\begin{figure}[ht]
\begin{center}
 \includegraphics[width=2.6in]{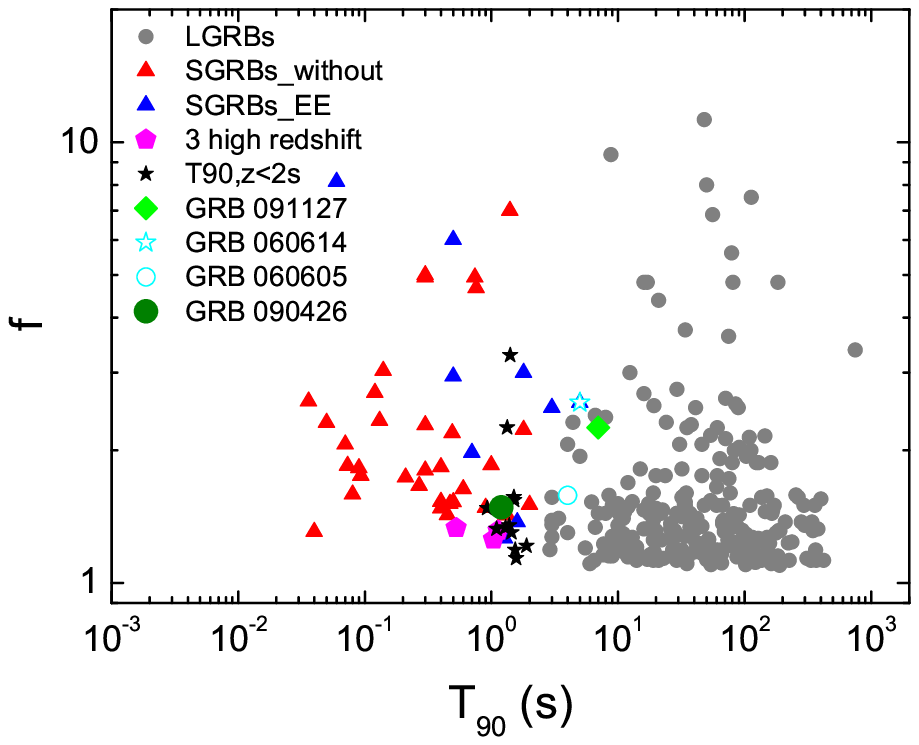} 
 \includegraphics[width=2.6in]{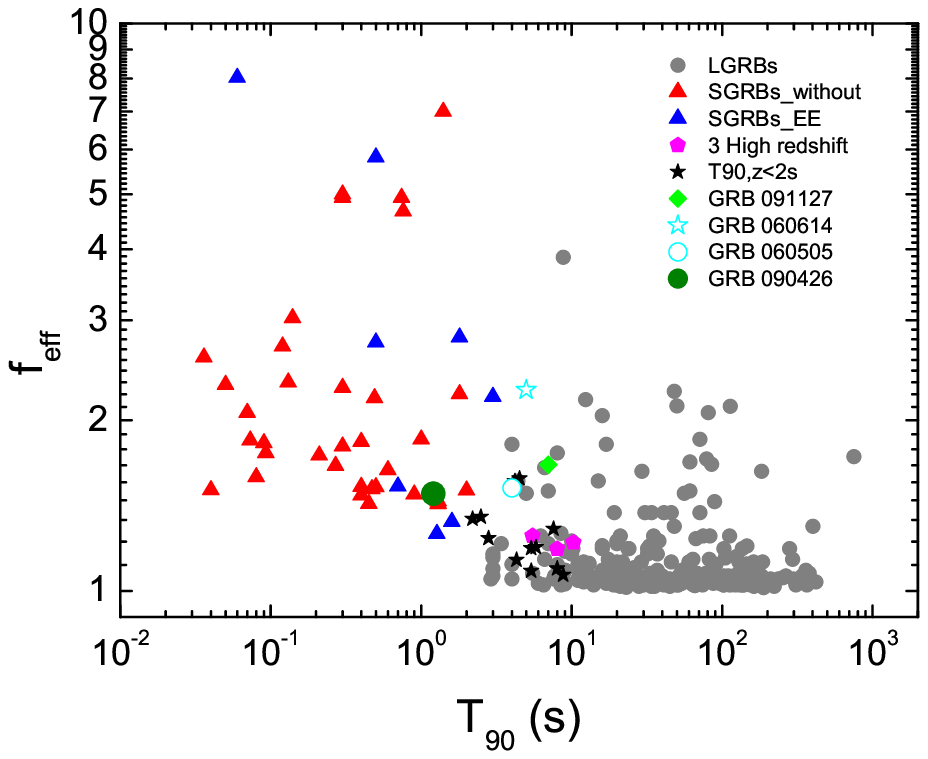} 
 \caption{{\em Left}: The $T_{90} - f$ distribution of long and short GRBs
in the Swift GRB sample. Both long and short GRBs can have high spikes with
large $f$. {\em Right}: The $T_{90} - f_{\rm eff}$ distribution of long 
and short GRBs. Short GRBs can have large $f$'s suggesting that they
are intrinsically short. However, some long GRBs can have $f_{\rm eff}$ 
reaching several. So short GRBs with low $f$ could be confused with 
``tip-of-iceberg'' long GRBs. An example is GRB 090426, which has
$f = 1.48$.  From H.-J. L\"u et al., 2012, in preparation.}
   \label{f}
\end{center}
\end{figure}

Such an exercise was carried out by L\"u et al. (2012).
Figure 4 shows the distribution of $f$ and $f_{\rm eff}$ for long and
short GRBs. The left panel shows both long and short GRBs can have 
high amplitude spikes. The right panel shows that if only the bright 
spike of a long GRB is above the background, the typical $f_{\rm eff}$
values are substantially dropped. Some short GRBs have
a large enough $f$ not confused with the bright tip of long GRBs.
This means that they are intrinsically short.
However, some short GRBs have an $f$ comparable to  
$f_{\rm eff}$ of some long GRBs, suggesting that they may be confused
with bright spike of long GRBs.  It is interesting to note that GRB 
090426 is a low $f$ short GRB ($f=1.48$), which falls into the 
``confusion'' area. Multiple observational criteria suggest
that it is a Type II GRB (e.g. Levesque et al. 2010; Xin et al. 2011).
So its short duration is likely due to the ``tip-of-iceberg'' effect.

One can also move a long GRB to progressively higher redshifts
until its rest-frame duration is shorter than 2 seconds. Defining
$f_{\rm eff,z} \equiv F'_{p,z}/F_b$, one can analyze the range of
$f_{\rm eff,z}$ and compare with the $f$ values of the observed
rest-frame short GRBs. It turns out that the three rest-frame short
GRBs at high redshifts (GRBs 080913, 090423, 090429B) all have $f$
falling into the confusion region, suggesting that they are consistent
with being the ``tip of iceberg'' of long-duration GRBs
(L\"u et al. 2012).


\vspace{0.3cm}
\noindent
{\bf Acknowledgments}\\
I thank my students/collaborators who carried out the heavy-duty
tasks to tackle the complicated problems discussed in this talk,
in particular, Hou-Jun L\"u, Francisco Virgili, Bin-Bin Zhang, and
En-Wei Liang. This work is supported by NSF through grant AST-0908362, 
and by NASA through grants NNX10AD48G, NNX10AP53G, and NNX11AQ08G.




\bibliographystyle{aipproc}   




\begin{discussion}

\discuss{N. Tominaga}{In your flowchart, a GRB is moved to Type I
if it has no SN component. It is not necessary to have a bright SN,
because there is a diversity in SN brightness.}

\discuss{B. Zhang}{First, the flowchart is based on what we see
not what we believe. Indeed some models can make faint SNe associated
with GRBs. However, observationally there is no robust evidence yet 
for massive star GRBs with no SN association. GRB 060614 is a long 
GRB. However, it is not that different from the ``smoking gun'' 
``short'' GRB 050724. GRB 050724 is sitting near the edge of an 
early type galaxy. However, it is
not that short, either. It has an early emission episode with 
$T_{90}\sim 3$ s, 
followed by an extended emission tail lasting longer than 100 s
(Barthelmy et al. 2005). GRB 060614, on the other hand, has an
early episode lasting for about $\sim 5$ s, followed by a softer
oscillating tail of $\sim 100$ s (Gehrels et al. 2006).
It would be essentially identical to GRB 050724 if it were 8 times
less energetic (Zhang et al. 2007). Its host galaxy does not have
active star formation, and the location of the burst does not track
the bright star forming region in the host galaxy. All these suggest
that GRB 060614 is not different from GRB 050724, and should be a 
Type I GRB. Interpreting it as a SN-less Type II GRB
is based on the {\em belief} that all long GRBs are associated with
massive stars. Otherwise, most nearby short GRBs (including the actually
long-duration ``short'' GRB 050724) do not have SN associations,
either. So why not interpreting them as SN-less core collapses as well?
Second, in my flowchart, a long GRB without a SN association is not
directly linked to Type I. It is only moved to the left hand side
(which becomes a Type I candidate). One needs to go through other
criteria to make the final judgment. GRB 060614 is eventually
grouped to Type I, but this is because its host galaxy and location
within the host galaxy also satisfy the criteria of a Type I GRB.
If, in the future, one detects a long duration, SN-less GRB
sitting in the bright region of an active star-forming galaxy,
having large energetics, and satisfying the empirical correlations 
of other Type II GRBs, one would move it back to the right hand side
of my flowchart, and it will be identified as a Type II candidate. 
If this indeed happens, then theoretical modeling of GRBs with 
very faint SNe becomes relevant.}

\discuss{F. Mirabel}{Can you exclude that Doppler boosting dependence
on viewing angle does not play any role in GRB classification? Namely,
on duration and hardness?}

\discuss{B. Zhang}{The Doppler effect plays an important role in
the AGN field, but less in the GRB field. This is mostly
because GRBs have higher Lorentz factors ($\Gamma > 100$) and wider jet 
opening angles ($\theta_j \sim 5^{\circ}$) than AGNs, so that usually 
$\Gamma^{-1} \ll \theta_j$ is satisfied. For most geometry, the line of
sight is within the jet cone. Observationally, long-duration GRBs
typically have multiple peaks, with minimum variability of order 
milliseconds. So the long duration is the intrinsic property of the
central engine, not due to Doppler broadening. Otherwise one should
not see much smaller variability time scales.
Also early optical afterglows of most GRBs show a decay 
behavior. This is consistent with the prediction of the on-beam geometry. 
If the line of sight is outside of the jet cone initially, one would 
expect to see a rising (or very shallow decay) lightcurve instead. Of
course, there is a category of low-luminosity, long-duration GRBs
that show one broad peak in the lightcurve. These bursts could be
related to bursts viewed at a large off-beam angle with a low Doppler
factor. A competitive model is that these are intrinsically different
events, probably related to jets that barely breakout from
the star, in contrast with high-luminosity long GRBs that have
successful jets. The two scenarios can be in principle differentiated
with late radio afterglow data. The current data do not support
the off-beam jet model. So, to answer your question, even though
the viewing angle effect is not fully excluded, one can say that
it plays a minor, if any, role in GRB classification.
}

\discuss{E. Nakar}{I have a few comments. First, the separation line 
of $T_{90} = 2$ s is for BATSE. For other detectors (e.g. Swift), the
separation line can be different. Second, with $\gamma$-ray information
only, it is impossible to tell which physical category a GRB belongs 
to. Besides contamination of Type II in short GRBs, there could be 
contamination of Type I in long GRBs, too. }

\discuss{B. Zhang}{I fully agree with both comments. Regarding 
$T_{90}$, I agree that the 2-second separation line is for BATSE
only. The Swift team currently also uses 2-second as the separation. 
On the other hand, as is described in the flowchart (Fig.1), the initial
separation is not fundamental. One really needs to go through all
the criteria before the final identification of the physical 
category of a burst is made. There are bridges to connect the
two sides. So even if there was inaccuracy in the initial duration
criterion, after going through the flowchart, a burst would land 
in the right physical category. Second, indeed there could be more
Type I contaminations in the long duration GRBs. For example, if
a long GRB (no short hard spike and no deep SN upper limit) sit
in the outskirt of an early type galaxy, one has to be cautious
to define its category. In the current flowchart, if this burst
also had low energetics, it would be grouped to the ``unknown''
category. This is a field full of surprises. The
flowchart was designed to our best knowledge in 2009, and it
still works reasonably well until now (see e.g. Kann et al.
2010, 2011). However, it is possible that future observations 
may suggest that more ``bridges'' are needed in the flowchart, or 
even the global structure of the flowchart has to be modified.}

\end{discussion}

\end{document}